\PassOptionsToPackage{draft,bookmarks=false}{hyperref}
\documentclass[conference]{IEEEtran}
\usepackage{algorithmic}
\usepackage{algorithm}
\usepackage{array}
\usepackage{amsmath}
\usepackage{booktabs}
\usepackage{caption}
\usepackage[caption=false,font=normalsize]{subfig}
\usepackage{textcomp}
\usepackage{stfloats}
\usepackage{url}
\usepackage{verbatim}
\usepackage{graphicx}
\usepackage[colorlinks=true, allcolors=blue]{hyperref} 
\usepackage{cite}
\usepackage{acronym} 
\usepackage{orcidlink} 
\acrodef{ZSM}{zero-touch network management}
\acrodef{LAM}{large AI model}
\acrodef{SLA}{service level agreement}
\acrodef{QES}{quality efficiency score}
\acrodef{CQI}{Composite Quality Index}
\acrodef{CoT}{Chain-of-Thought}
\acrodef{RL}{reinforcement learning}
\acrodef{AI}{artificial intelligence}
\acrodef{AI/ML}{artificial intelligence \& machine learning}
\acrodef{ANN}{artificial neural network}
\acrodef{ASE}{amplified spontaneous emissions}
\acrodef{CUT}{channel under test}
\acrodef{DRL}{deep reinforcement learning}
\acrodef{EGN}{enhanced Gaussian noise}
\acrodef{EON}{elastic optical network}
\acrodef{GT}{ground truth}
\acrodef{GSNR}{generalized signal-to-noise ratio}
\acrodef{KDE}{kernel density estimation}
\acrodef{LLM}{large language model}
\acrodef{LP}{lightpath}
\acrodef{MAE}{mean absolute error}
\acrodef{MF}{modulation format}
\acrodef{ML}{machine learning}
\acrodef{NLI}{non-linear impairments}
\acrodef{QoT}{quality of transmission}
\acrodef{RMSA}{routing, modulation and spectrum assignment}
\acrodef{XGB}{XGBoost}
\acrodef{IBN}{intent-based networking}
\acrodef{BER}{bit error rate}
\acrodef{XAI}{explainable artificial intelligence}
\acrodef{ADON}{autonomous driving optical network}
\acrodef{SHAP}{SHapley Additive exPlanations}
\IEEEoverridecommandlockouts

\begin{document}
\pagestyle{empty}

\title{Human‑Grounded Evaluation of Large Language Models for Optical Network Automation}

\author{
    \IEEEauthorblockN{Kiarash Rezaei\IEEEauthorrefmark{1}\orcidlink{0009-0003-6166-2614}, 
    Omran Ayoub\IEEEauthorrefmark{2}\orcidlink{0000-0002-3884-3594},
    Paolo Monti\IEEEauthorrefmark{1}\orcidlink{0000-0002-5636-9910}, Carlos Natalino\IEEEauthorrefmark{1}\orcidlink{0000-0001-7501-5547}}
   \\ \IEEEauthorblockA{\IEEEauthorrefmark{1} Department of Electrical Engineering, Chalmers University of Technology, 412 96 Gothenburg, Sweden
    \\ \{kiarashr, mpaolo, carlos.natalino\}@chalmers.se} \\
    \IEEEauthorblockA{\IEEEauthorrefmark{2}University of Applied Sciences and Arts of Southern Switzerland, 6928 Lugano, Switzerland
    \\ omran.ayoub@supsi.ch}
    }

\maketitle
\IEEEpubidadjcol
\begin{abstract}
\Acp{LLM} are increasingly adopted for network automation, yet their output quality and inference cost can vary substantially across \ac{LLM} families.
We present HuGLEN, a stepwise evaluation pipeline that uses an \ac{LLM}-as-a-judge together with a small set of expert ratings to enable scalable and reproducible comparison of candidate \acp{LLM}, and to rank them using a \ac{QES}.
We demonstrate HuGLEN for translating outputs from an \ac{XAI} model for the optical network \ac{QoT} estimation task into operator-friendly explanations.
Our results show that a medium-sized \ac{LLM} (12B parameters) achieves the highest \ac{QES}, indicating the best trade-off between explanation quality and efficiency.
Overall, HuGLEN reduces the human-labeling burden while supporting consistent model selection for operator-facing automation tasks.
\end{abstract}


\acresetall
\section{Introduction}
\Acp{LLM} are transforming network automation by enabling unprecedented capabilities in configuration generation, fault diagnosis, and operational decision-making.
From automating complex network configurations that traditionally required specialized expertise to providing natural language interfaces for network management, \acp{LLM} demonstrate remarkable potential across diverse networking domains.

Recent research has demonstrated \ac{LLM} applications across networking domains, including optical network control and lifecycle management~\cite{Song25,Liu:25} and wireless network design and optimization~\cite{qiu2024,wu2025deepformreasoninglargelanguage}.
Beyond domain-specific applications, the broader networking community has embraced \ac{LLM}-based automation tools, including benchmark frameworks for network-configuration tasks such as NetConfEval~\cite{wangNetConfEval2024}.
In parallel, telecom-focused evaluation resources and benchmarks are emerging to better characterize \acp{LLM} performance under domain-specific requirements~\cite{lee2024telbench}.
Related work has also integrated \acp{LLM} with XAI-enabled anomaly detection for zero-touch service management (ZSM) in 6G microservices, using the \ac{LLM} to translate numerical \ac{XAI} outputs into human-readable explanations and to autonomously execute corrective actions for service level agreement (SLA) violations~\cite{mekrache2024}.

However, this rapid adoption presents critical challenges.
The performance of \acp{LLM} vary widely across model families and task types, their energy consumption can be substantial, and reliable evaluation methods that align with human judgment remain elusive.
While deploying state-of-the-art \acp{LLM} may appear attractive, doing so for routine network tasks often leads to unnecessary computational overhead and energy expenditure, especially when smaller, more efficient ones could deliver comparable performance.
Recent benchmarking results indicate that some \acp{LLM} may consume over 70 times more energy per query than streamlined alternatives~\cite{jegham2025hungryaibenchmarkingenergy,fernandez2025energyinference}, underscoring the need for evaluation frameworks that balance output quality with computational efficiency.
Existing evaluation methods offer only partial solutions: text-similarity metrics (e.g., BLEU/ROUGE) capture surface overlap but miss deeper dimensions like correctness and clarity~\cite{PapineniRWZ02,lin_rouge_2004}, while human evaluations are informative but costly and difficult to scale~\cite{Ayoub_2025_icton,RezaeiWiMob2025}.
Together, these limitations reveal a fundamental gap: the absence of a scalable and trustworthy way to evaluate \acp{LLM} that can both reflect human judgment and account for computational efficiency.
Bridging this gap is essential to guide the resource-aware adoption of \acp{LLM} in network automation. 

One approach to address this challenge is LLM-as-a-judge, where an \ac{LLM} evaluates the outputs of other \acp{LLM} using explicit, task-specific rubrics~\cite{zheng2023judgingllmasajudgemtbenchchatbot, maatouk2025telellmsseriesspecializedlarge, liu2023geval, pmlr-v235-chiang24b}. This automation significantly reduces the reliance on extensive human annotation, thereby improving scalability and mitigating individual cognitive bias. However, prior work has shown that \acp{LLM} employed as evaluators may exhibit reduced agreement with subject-matter experts in tasks requiring specialized domain knowledge. This limitation underscores the importance of careful rubric design and human involvement, particularly in technical domains such as telecommunications~\cite{szymanski2025limitations}.

\newpage
To this end, this paper introduces \textbf{HuGLEN} (Human-Grounded Auto LLM Evaluation), a modular framework that bridges scalable LLM judge evaluation with human expert involvement in the assessment of \acp{LLM}, with a particular focus on network automation scenarios and use cases where the quality of \ac{LLM} outputs directly impacts operational decisions.
In this work, we use standardized prompts (instead of task-specific fine-tuning) because task-specific training data are often unavailable in operational networks
\footnote{To support reproducibility, we will release the code, prompts, and evaluation artifacts upon acceptance.}.

HuGLEN employs a small set of human-based evaluations to establish a baseline for \ac{LLM} performance, which is then leveraged to calibrate an automated evaluation pipeline. 
This automation significantly reduces the reliance on extensive human annotation, making the evaluation process more scalable and less prone to individual cognitive bias. 
To guide resource-aware deployment and adoption, HuGLEN introduces the \acf{QES} concept, a novel metric that jointly considers the quality of \ac{LLM}-generated outputs and computational efficiency. 
Jointly, the innovations proposed in HuGLEN enable more efficient and consistent selection of \acp{LLM} for network automation tasks, helping to prevent unnecessary resource consumption while ensuring that \ac{LLM} choices remain grounded in human-relevant criteria.
We evaluate the proposed framework in a representative use case, i.e., assessing \acp{LLM} used to generate natural language explanations for a \ac{QoT} estimation model in optical networks \cite{RezaeiWiMob2025}.
Results show that a mid-sized \ac{LLM} (12B parameters) achieves the highest \ac{QES}, indicating the best trade-off between explanation quality and computational efficiency for this use case.

\section{HuGLEN: Human-grounded Auto LLM Evaluation}
\label{sec_proposed_framework}

\begin{figure}[t]
   \centering
   \includegraphics[width=\columnwidth]{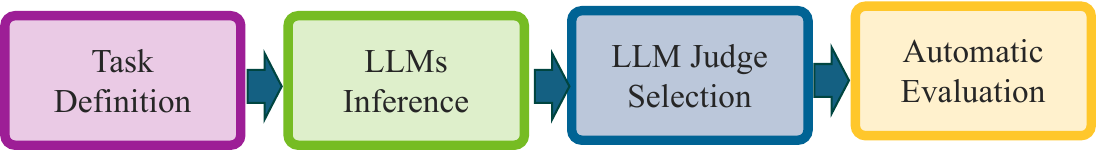}
   \caption{High-level overview of HuGLEN Framework.}
   \label{fig:HuGLEN}
\end{figure}

This section describes the HuGLEN workflow (Fig.~\ref{fig:HuGLEN}).
HuGLEN takes as input task data and a task specification, which are converted into a set of standardized prompts via a prompt generation module.
The workflow then proceeds through four stages: \textit{(i) LLMs Inference}, where candidate \acp{LLM} generate outputs and the human-centered evaluation metrics are defined; \textit{(ii) \ac{LLM} Judge Selection}, where candidate \ac{LLM} judges are compared against expert ratings using agreement scores to select a high-agreement \ac{LLM} judge; and \textit{(iii) Automatic Evaluation}, where the selected \ac{LLM} judge scores all candidate outputs and the \ac{QES} is computed to rank the \acp{LLM}.

While task-specific adaptation (e.g., instruction tuning or LoRA fine-tuning) may further improve performance~\cite{hu2021lora,xin2024lora_instruction}, it typically requires collecting and curating task data, which can be difficult in operational networks due to data access constraints (e.g., proprietary datasets, privacy), or limited labeling capacity.
Therefore, in this work we deliberately focus on prompt engineering and standardized prompting (rather than fine-tuning).

\begin{figure}[!t]
  \centering
  \begin{minipage}{0.99\linewidth}
    \includegraphics[width=\linewidth]{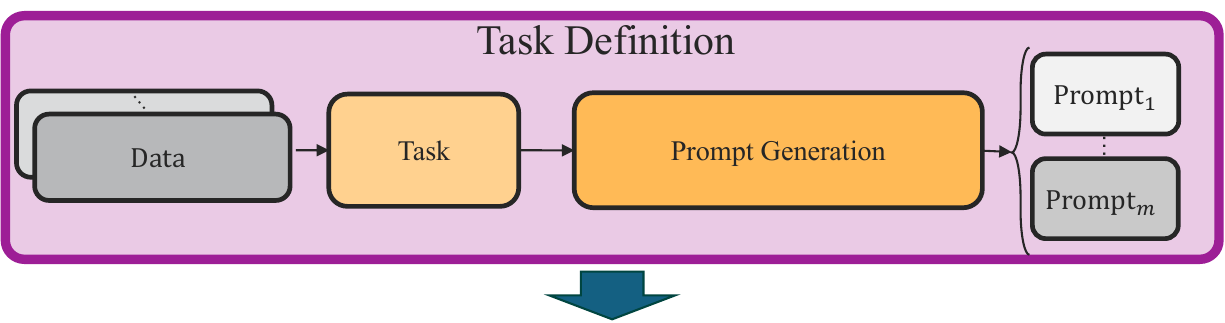}
    \vspace{-6.55mm}
    \hspace{0mm}
  \end{minipage}\vspace{2.2mm}
  \begin{minipage}{0.99\linewidth}
    \includegraphics[width=\linewidth]{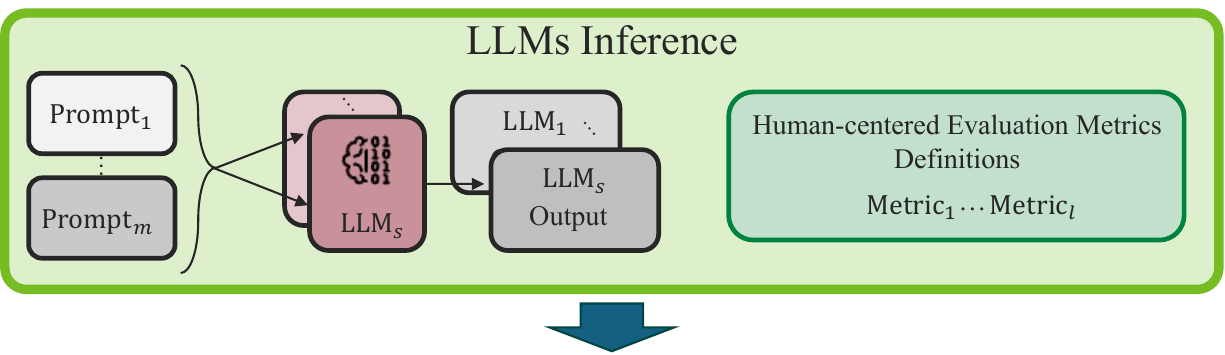}
    \vspace{-6.55mm}
    \hspace{0mm}
  \end{minipage}\vspace{2.2mm}
  \begin{minipage}{0.99\linewidth}
    \includegraphics[width=\linewidth]{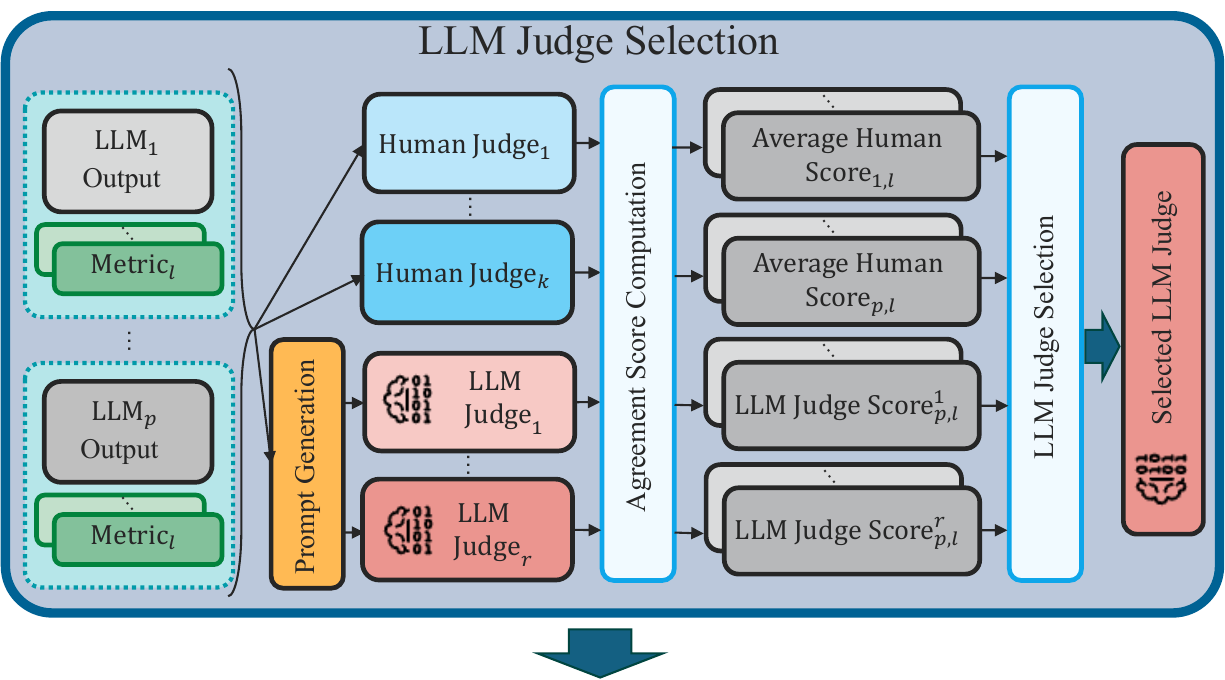}
    \vspace{-6.55mm}
    \hspace{0mm}
  \end{minipage}\vspace{2.2mm}
  \begin{minipage}{0.99\linewidth}
    \includegraphics[width=\linewidth]{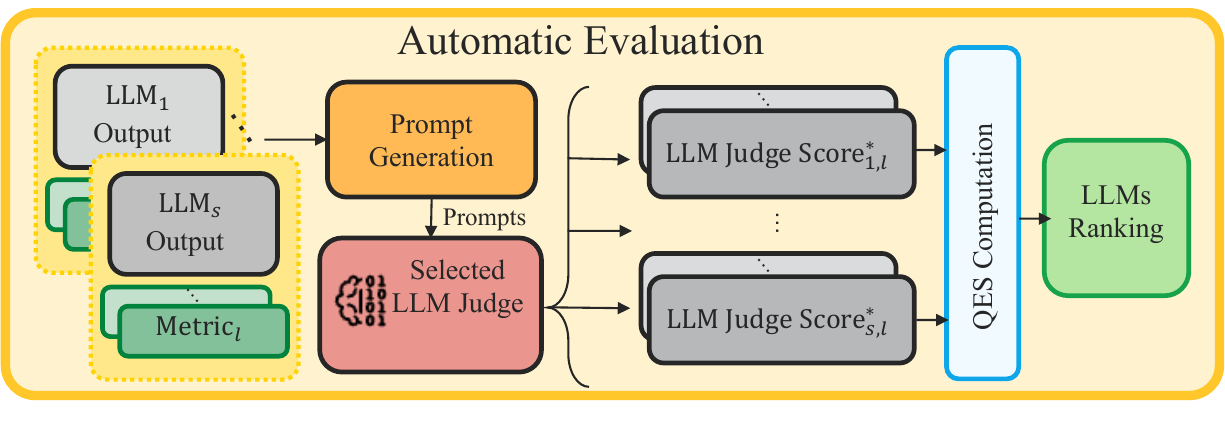}
    \vspace{-4mm}
    \hspace{0mm}
  \end{minipage}
    \caption{
    Detailed workflow of the HuGLEN framework, expanding on the high-level overview in Fig.~\ref{fig:HuGLEN}. 
    The diagram illustrates each sequential stage: The task definition formalizes the transformation of input data into structured prompts via the prompt generation module, \ac{LLM} inference including output generation and definition of human-centered evaluation metrics, LLM judge selection based on agreement with human ratings, and automated evaluation with \ac{QES}-based ranking of candidate models.
}
  \label{fig:HuGLEN_framework}
\end{figure}

\subsection{Task Definition}

The \textit{Task Definition} step specifies the target task, inputs, and a standardized prompting template (via a prompt generation module) to ensure all candidate \acp{LLM} are evaluated under identical conditions.
The framework itself is domain- and task-agnostic.
It can be adapted to any application that leverages an \ac{LLM}, with the prompt-generation module tailored to the specific objective.

As illustrated in the first (purple) block of Fig.~\ref{fig:HuGLEN_framework}, \textit{Task Definition} maps available \emph{data} to a concrete \emph{task} specification and then uses the prompt generation module to convert the task inputs into a set of standardized prompts.

\subsection{LLMs Inference}
\label{subsec:llms inference}

In \textit{LLMs Inference}, second (green) block in Fig.~\ref{fig:HuGLEN}, all candidate \acp{LLM} generate outputs for the same set of prompts.
We also define a small set of human-centered objective metrics, used consistently by both human evaluators and candidate \ac{LLM} judges.
\subsection{LLM Judge Selection}

In the third stage, HuGLEN identifies the most reliable automated evaluator, referred to as the \emph{\ac{LLM} judge}, to score the outputs of the candidate \acp{LLM}.
Here, ``reliable'' refers to the \ac{LLM} judge's ability to produce scores that show high agreement with human judgments and remain consistent across different input samples.
As illustrated in the third (blue) block of Fig.~\ref{fig:HuGLEN_framework}, the process begins with the set of outputs generated by the candidate \acp{LLM} in the previous stage.
These outputs are first evaluated by multiple human judges using the human-centered metrics defined in Section~\ref{subsec:llms inference}. 

For each candidate \ac{LLM}, the scores assigned by all human judges are averaged across each evaluation metric, yielding a human reference score per candidate \acp{LLM} and per metric.
In parallel, the prompt generation module produces standardized prompts for a set of candidate \ac{LLM} judges, which evaluate the same outputs using the same metrics.
Similarly, for each \ac{LLM} judge, the results are recorded separately for every candidate \ac{LLM} and for each metric.
This setup enables a direct comparison between human and automated evaluations, providing a basis for selecting the most reliable \ac{LLM} judge.
The \ac{LLM} judge that demonstrates the highest overall alignment with human evaluations is selected to perform the scoring in the final automatic evaluation stage.
Alignment can be quantified using different approaches, such as inter-rater agreement measures.
This procedure have the potential to maximize consistency with human judgment while substantially reducing the need for manual evaluation.
The achieved level of alignment is reported to indicate the reliability of the selected \ac{LLM} judge.

\subsection{Automatic Evaluation}
\label{subsec:automaic_evaluation}

The final stage of HuGLEN involves automated scoring of candidate \acp{LLM} using the selected \ac{LLM} judge.
As shown in the  last (yellow) block of Fig.~\ref{fig:HuGLEN_framework}, the \ac{LLM} judge evaluates each candidate \ac{LLM}'s outputs based on the human-centered metrics introduced in Section~\ref{subsec:llms inference}. 
These individual metric scores are then aggregated into a single weighted average, yielding an overall quality score for each candidate \ac{LLM}.
By default, HuGLEN assigns equal weights to each metric, but practitioners can customize these weights to emphasize specific evaluation criteria according to their operational priorities.

\subsubsection{Composite Quality Index (CQI)}
We define the \acf{CQI} as an aggregate quality metric for model $i$, combining correctness $C_i\in[0,1]$, scope $S_i\in[0,1]$, and usefulness $U_i\in[0,5]$.
We first normalize usefulness as $\widehat{U}_i = U_i/5 \in [0,1]$, and then compute
\begin{equation}
\mathrm{CQI}_i = \frac{w_1\,C_i + w_2\,S_i + w_3\,\widehat{U}_i}{w_1+w_2+w_3}.
\end{equation}
By default, $w_1=w_2=w_3=1$, hence $\mathrm{CQI}_i=(C_i+S_i+\widehat{U}_i)/3$.

\subsubsection{Quality-Efficiency Score (QES)}
To jointly rank models by quality and efficiency, we normalize the \ac{CQI} values across the evaluated model set:
\begin{equation}
\widehat{\mathrm{CQI}}_i = \frac{\mathrm{CQI}_i - \min_j(\mathrm{CQI}_j)}{\max_j(\mathrm{CQI}_j) - \min_j(\mathrm{CQI}_j)}.
\end{equation}
We then define an efficiency factor based on parameter count $P_i$ as a proxy for inference cost, relative to the largest model $P_{\max}$:
\begin{equation}
\eta_i = \frac{P_{\max}}{P_i},
\end{equation}
and normalize it to $[0,1]$ using the smallest model $P_{\min}$:
\begin{equation}
\widehat{\eta}_i = \frac{\eta_i - 1}{\frac{P_{\max}}{P_{\min}} - 1}.
\end{equation}
Finally, we compute
\begin{equation}
\mathrm{QES}_i = 100\,\Big(\alpha\,\widehat{\mathrm{CQI}}_i + (1-\alpha)\,\widehat{\eta}_i\Big),
\end{equation}
where $\alpha\in[0,1]$ controls the quality--efficiency trade-off (default $\alpha=0.7$). In deployments, the proxy $\widehat{\eta}_i$ can be replaced by measured latency, energy, or monetary cost.

\begin{figure*}[!t]
    \centering
    \subfloat[Agreement of correctness]{\includegraphics[width=0.33\linewidth]{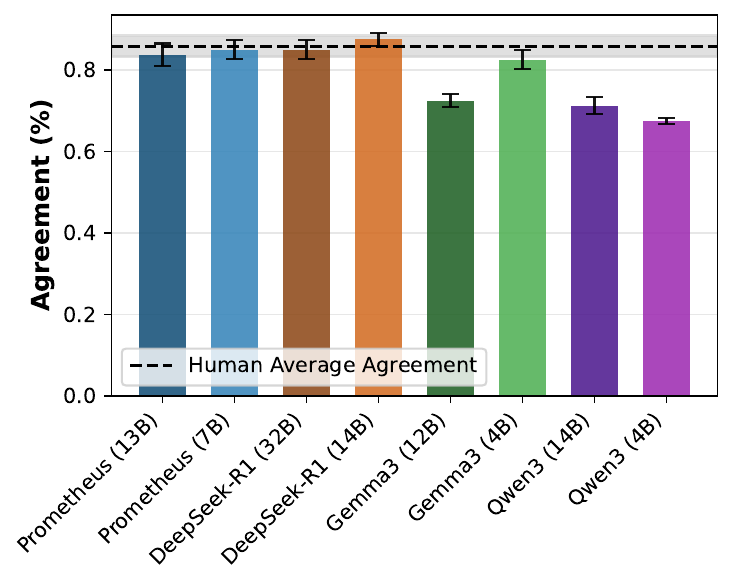}}
    \hfill
    \subfloat[Agreement of scope]{\includegraphics[width=0.33\linewidth]{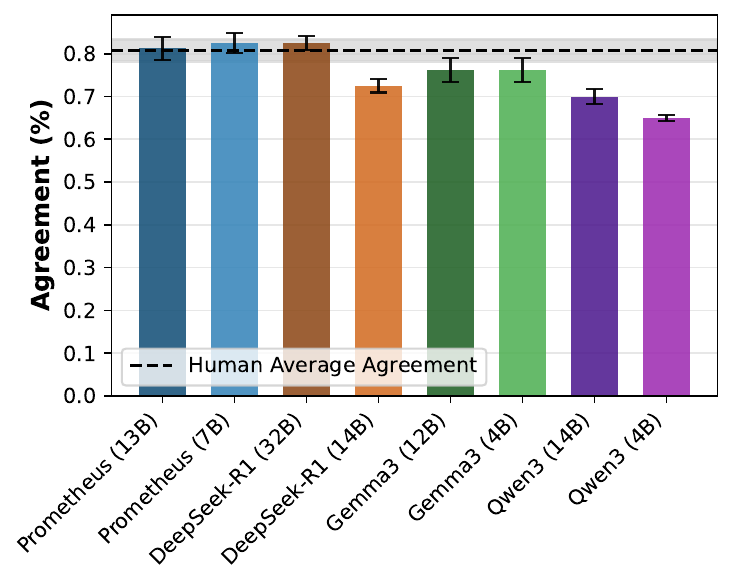}}
    \hfill
    \subfloat[Agreement of usefulness]{\includegraphics[width=0.33\linewidth]{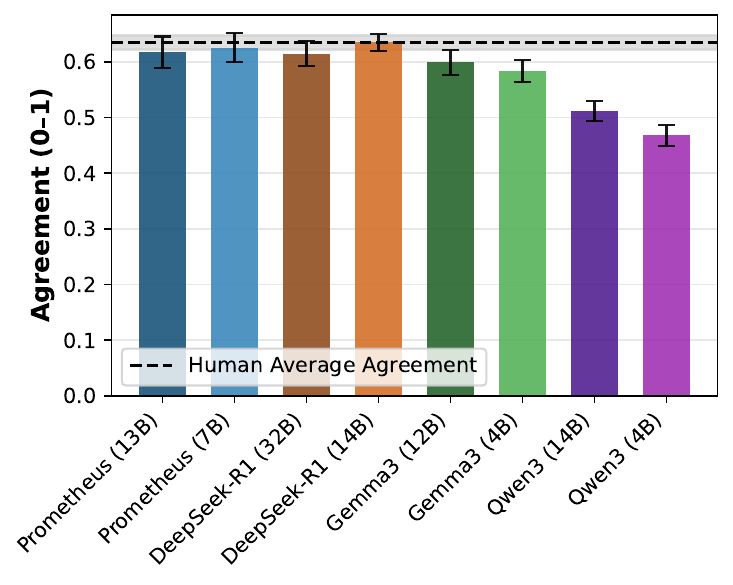}}
        
    \caption{
Agreement scores between candidate \ac{LLM} judges and human evaluators across the three predefined metrics for QoT explanation assessment: (a) correctness, (b) scope, and (c) usefulness. Each plot shows the level of alignment for each LLM judge and average human reference across 40 explanation samples.
}
    \label{fig:three_figures}
\end{figure*}

\section{Use Case: AI-based Optical \ac{QoT} Estimation}
\label{sec:usecase}
We illustrate HuGLEN on a representative network-automation scenario: \ac{AI}-based \ac{QoT} estimation for unestablished lightpaths in optical networks.
In this setting, the estimated \ac{QoT} informs resource-allocation decisions that are supervised by an optical network engineer who is typically not an \ac{AI} specialist.

\begin{figure}[t]
    \centering
    \includegraphics[width=\columnwidth]{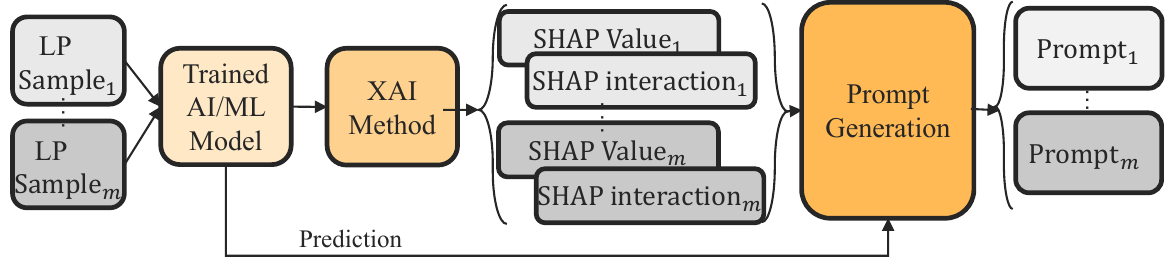}
    \caption{Use case workflow (Task Definition stage).}
    \label{fig:use_case}
\end{figure}

When a lightpath decision is flagged as unusual (e.g., the predicted \ac{QoT} contradicts operational expectations), the engineer must understand \emph{why} the \ac{AI} model produced that output.
A common starting point is a SHAP-based local explanation, consisting of SHAP values (local feature importance) and SHAP feature-interaction plots (joint feature effects on the \ac{ML} prediction), as illustrated in Fig.~\ref{fig:shap_results}.
However, these numerical attributions are difficult to interpret without \ac{AI} expertise and may be prone to cognitive bias.
We therefore use an \ac{LLM} to translate the SHAP values into operator-facing, human-readable, contextualized explanations.
Our expectation is that such contextualized explanations, when analyzed in conjunction with the SHAP values and interactions, will require a level of expertise from the engineer, and result in a better understanding of the predictions. 
The translation task just described corresponds to the \emph{Task Definition} stage of HuGLEN.
As illustrated in Fig.~\ref{fig:use_case}, SHAP values, SHAP interactions, and predictions associated with each \ac{LP} sample are integrated into a structured prompt through the prompt generation module.
These structured prompts serve as inputs for the subsequent stages of HuGLEN.

\section{Experimental and Empirical Results}
\label{sec:experimental}

Building on the use case in Section~\ref{sec:usecase}, we instantiate the \textit{Task Definition} stage in Fig.~\ref{fig:HuGLEN_framework} for \ac{QoT} explanation generation.
For this purpose, we selected a set of candidate \acp{LLM}, including DeepSeek-R1 (1.5B, 14B, 32B), Gemma3 (4B, 12B), and Qwen3 (4B, 14B).

\begin{figure}[!t]
    \centering
    \subfloat[SHAP values]{\includegraphics[width=0.5\linewidth]{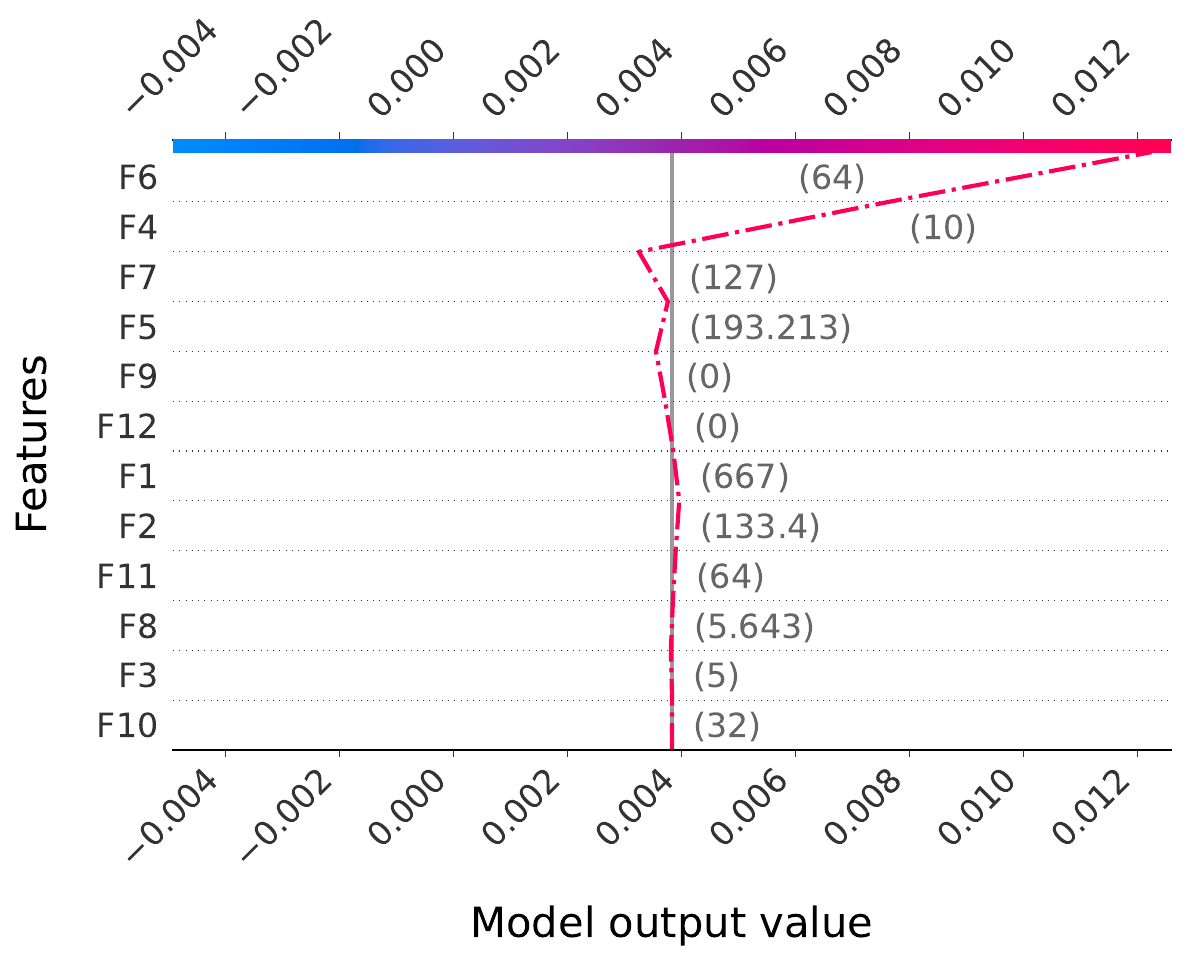}}
    \hfill
    \subfloat[SHAP interaction]{\includegraphics[width=0.5\linewidth]{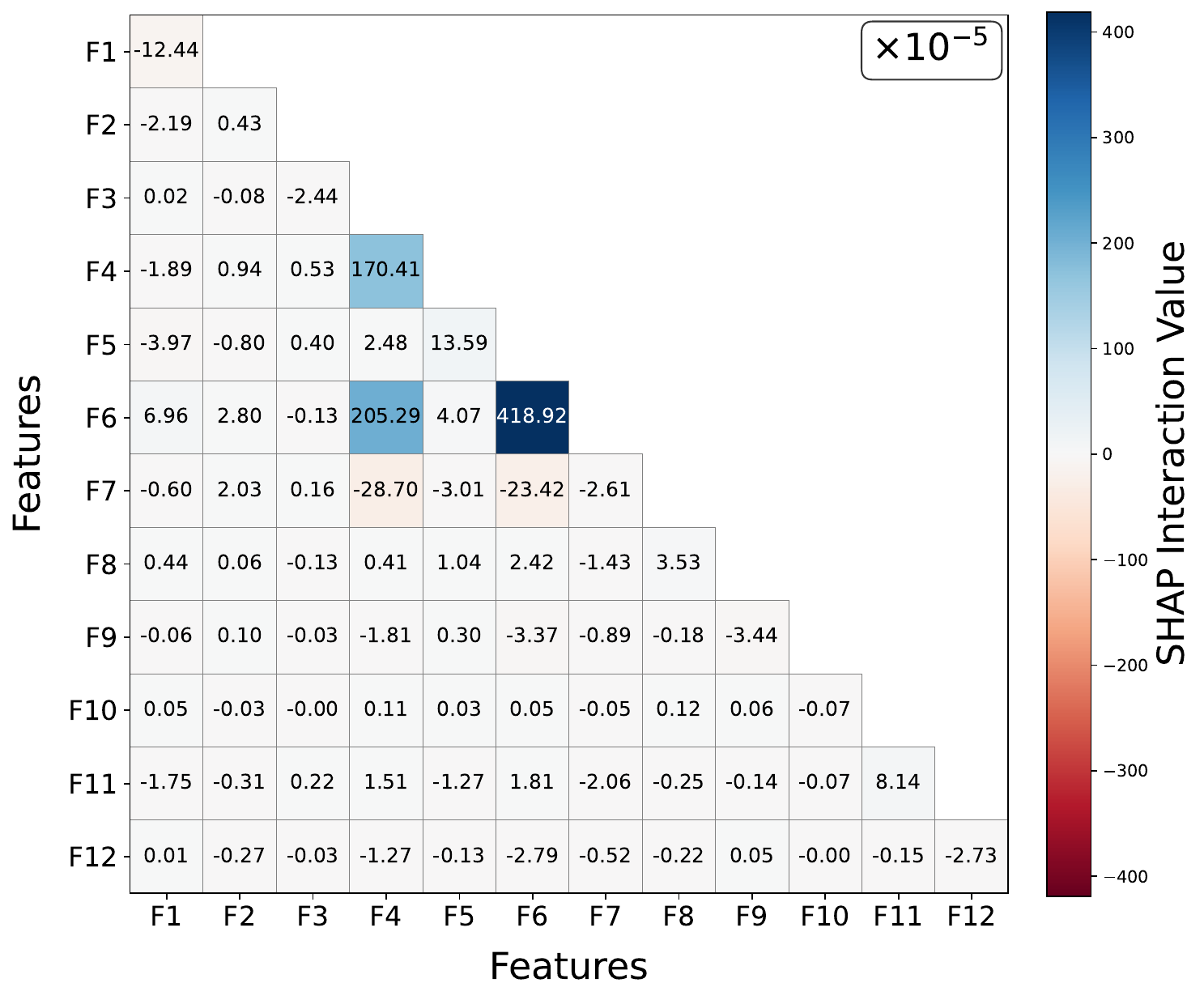}}
    \caption{An example of SHAP values and SHAP interaction plots, representing a local explanation generated by the \ac{XAI} method for one of the sample lightpaths whose \ac{QoT} was predicted by the \ac{AI/ML} model.}    \label{fig:shap_results}
\end{figure}

The process begins with training an \ac{XGB} regressor model to predict the bit error rate (BER) for various lightpaths using a set of representative features from data samples in \cite{bergk2021ml}.
Once the model is trained, we apply \ac{SHAP} as the \ac{XAI} method~\cite{lundberg_unified_2017}, producing SHAP values (per-feature contributions) and SHAP interaction values (pairwise feature effects), as illustrated in Fig.~\ref{fig:shap_results}.
We focus on local explanations to support case-by-case analysis of individual lightpaths.
The resulting SHAP outputs are embedded (by the prompt generation module) into structured prompts that are provided to candidate \acp{LLM}, enabling a direct comparison of their ability to produce operator-facing explanations.

In the \textit{LLMs Inference} phase, all candidate \acp{LLM} were prompted with the structured prompts generated in the previous stage, aiming at translating the SHAP values and interaction values into natural language explanations.
In terms of human-centered evaluation metrics, we rely on the following metrics \cite{Ayoub_2025_icton}:
\textit{(i) correctness}, measuring how faithfully interpretations reflect the underlying model explanation;
\textit{(ii) scope}, determining if correct interpretations emphasize the most critical aspects; and
\textit{(iii) usefulness}, gauging how effectively explanations assist human understanding.
Correctness and scope were assessed as binary outcomes, while usefulness was rated numerically from 0 (not helpful) to 5 (extremely helpful). 

During the \textit{LLM Judge Selection} stage, we followed a structured process to identify the most reliable \ac{LLM} judge for automated evaluation.
First, we selected DeepSeek-R1 (32B) as the representative \ac{LLM} from the previous phase and randomly extracted 40 natural language explanations.
To minimize labeling effort and ensure a consistent and high-quality benchmark, we relied on a single \ac{LLM} for this step.

Four human experts independently evaluated the explanations using the three previously defined human-centered metrics (correctness, scope, and usefulness), producing the baseline scores against which we later compared the LLM judges' assessments.
Additionally, the experts conducted their evaluations individually and without access to each other's assessments, avoiding potential sources of bias.
All participants provided informed consent for the use of their anonymized evaluations in this study.

Next, we considered a set of candidate \ac{LLM} judges, including Prometheus (7B-v2.0, 13B-v1.0), DeepSeek-R1 (14B, 32B), Gemma3 (4B, 12B), and Qwen3 (4B, 14B).
Prometheus models were specifically designed for evaluation and benchmarking tasks \cite{kim2024prometheus}.

Then, using the carefully designed prompt generation module, candidate \ac{LLM} judges were given task descriptions and the extracted explanations and asked to rate them for correctness, scope, and usefulness.
Their output evaluations were then compared against the human baseline, using agreement score as the primary selection criterion.
For correctness and scope, agreement was computed as the percentage of matching ratings between the \ac{LLM} judge and human evaluators.
For usefulness, we measured the standard deviation of paired ratings, inverted and normalized it to a 0–1 scale so that higher values represent closer alignment \cite{thakur-etal-2025-judging}.

Results showed that Prometheus-7B-v2.0 achieved the highest agreement scores on all three metrics, with Prometheus-13B-v1.0 close behind.
These two models were the top performers across metrics, as illustrated in Fig.~\ref{fig:three_figures} (a)--(c). DeepSeek-32B fell slightly behind in usefulness agreement \big(Fig.~\ref{fig:three_figures} (c)\big), whereas DeepSeek-14B showed a noticeable drop in scope agreement \big(Fig.~\ref{fig:three_figures} (b)\big) despite competitive correctness \big(Fig.~\ref{fig:three_figures} (a)\big).
Finally, based on its balanced and robust agreement with human ratings, Prometheus-7B-v2.0 was selected as the final \ac{LLM} judge for the automated evaluation stage.

\begin{figure}[t]
   \centering
    \includegraphics[width=\linewidth]{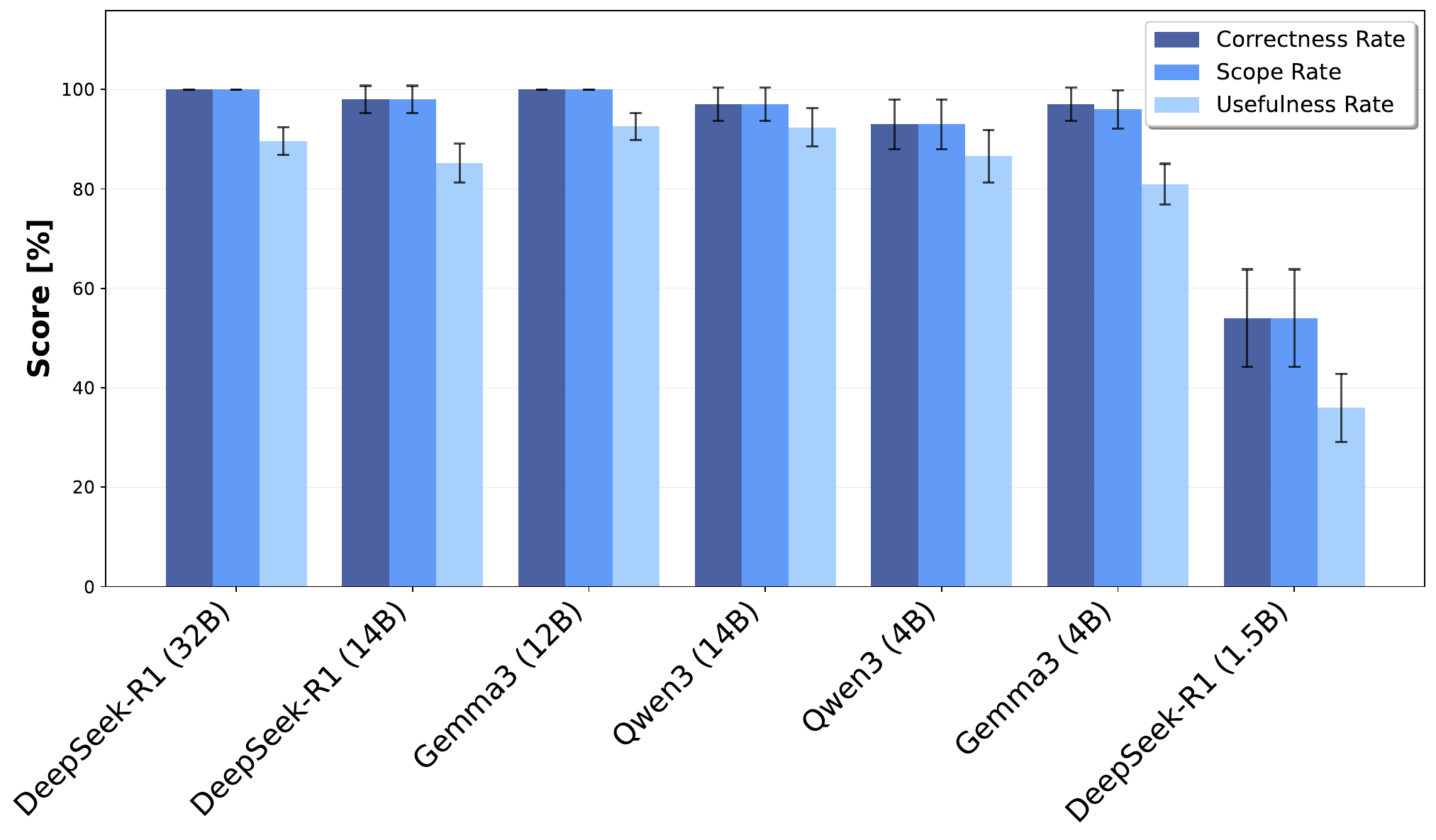} 
    \caption{
Comparative performance analysis of candidate \acp{LLM} evaluated by Prometheus-7B-v2.0 across correctness, scope, and usefulness rates.}
    \label{fig:performance_analysis}
\end{figure}

In the \textit{Automatic Evaluation} phase, we leveraged the selected \ac{LLM} judge (Prometheus-7B-v2.0) to automatically score 100 natural language explanation outputs from each candidate \ac{LLM}. 
Fig.~\ref{fig:performance_analysis} summarizes the quality of the natural language explanations produced by \acp{LLM}, which are automatically evaluated by the \ac{LLM} judge (Prometheus-7B-v2.0), across correctness, scope, and usefulness. Results show that most \acp{LLM} achieved near-optimal and very similar results for correctness and scope, reflecting the generally strong performance of all candidate \acp{LLM} on these metrics.
In contrast, the usefulness scores showed a wider spread, providing a clearer basis for differentiating between \acp{LLM}.

\begin{figure}[t]
   \centering
    \includegraphics[width=\linewidth]{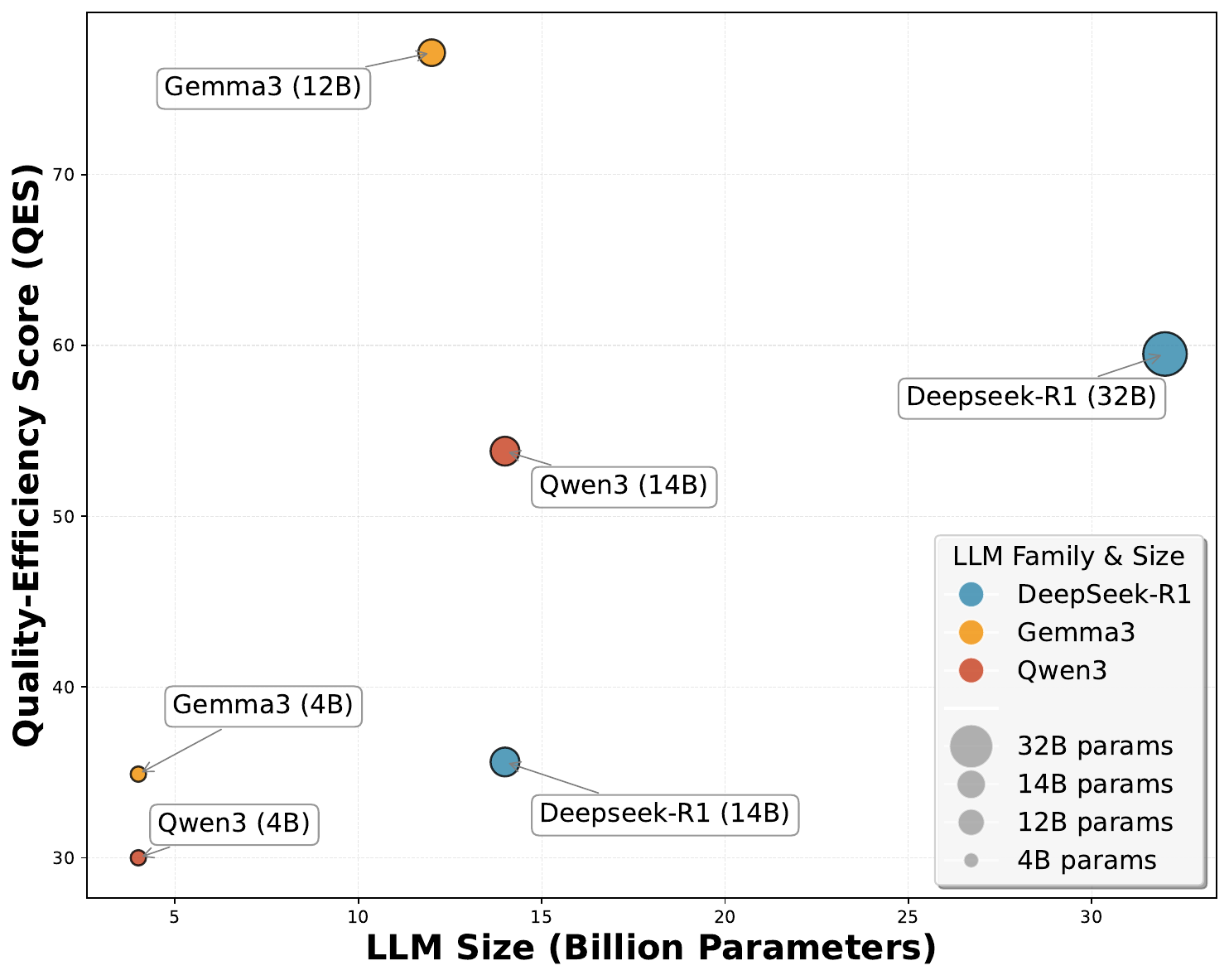} 
    \caption{
\Acf{QES} comparison for candidate LLMs across different number of parameters.}
    \label{fig:quality_efficiency}
\end{figure}

Finally, to select the \ac{LLM} that best balances explanation quality and computational efficiency, we computed \ac{QES}.
For this task, explanation quality is considered more important than efficiency, and thus we adopted the default weighting scheme mentioned in Section \ref{subsec:automaic_evaluation}.
Fig.~\ref{fig:quality_efficiency} shows the computed \ac{QES} for all candidate \acp{LLM} alongside their number of parameters.
Gemma3 (12B) achieved the highest \ac{QES}, at approximately 78 units, representing the best trade-off between quality and efficiency.
Interestingly, DeepSeek-R1 (32B) and Qwen3 (14B) achieved slightly lower scores despite their larger sizes, suggesting diminishing returns in quality relative to their computational cost.
In contrast, smaller \acp{LLM} such as Gemma3 (4B) and Qwen3 (4B) scored roughly 35--34 units lower, highlighting that their computational advantages come at the cost of substantially reduced output quality.
Overall, the results indicate that mid-sized \acp{LLM} (10--15B parameters) can offer the most attractive balance between explanation quality and efficiency for this task.

\section{Conclusion}

In this paper, we introduced HuGLEN, a novel framework for evaluating \acfp{LLM} intended for network automation tasks.
HuGLEN enables the automatic selection of the most trustworthy and resource-efficient \ac{LLM} for a given task, leveraging human-centered evaluation metrics to ensure both quality and practical relevance.
The proposed \acf{QES} allows to balance the \ac{LLM} output quality with computational efficiency and promote sustainable, evidence-driven deployment decisions.

We demonstrated HuGLEN’s effectiveness in a practical scenario, where the goal was to generate natural language explanations based on XAI interpretations of \acf{QoT} predictions in optical networks.
In this setting, \acp{LLM} translated the technical outputs of \acf{ML} models into clear, operator-friendly explanations that help network engineers better understand and trust automated decisions.

Experimental results showed that our approach can identify \acp{LLM} that provide strong explanatory performance while minimizing resource consumption.
Notably, Gemma3 (12B) was found to deliver the best trade-off between quality and efficiency, as measured by the \acf{QES}, while our automated \ac{LLM} judge selection process reduced the need for extensive manual evaluation.

\section*{Acknowledgements}
This work was supported by the Celtic-Next Flagship SUSTAINET-Advance project funded in Sweden by VINNOVA (2025-02987) and in Switzerland by Innosuisse (No. 119.588 INT-ICT).
The computations were enabled by resources provided by the National Academic Infrastructure for Supercomputing in Sweden (NAISS), partially funded by the Swedish Research Council through grant agreement no. 2022-06725.

\IEEEtriggeratref{11}
\bibliographystyle{IEEEtran}
\bibliography{references}

\end{document}